\documentclass[%
reprint,
superscriptaddress,
amsmath,amssymb,
floatfix,
aps,
pra,
]{revtex4-1}

\usepackage{float}
\usepackage{graphicx}   
\graphicspath{{../}}   %add figure path

\usepackage{epstopdf}
\usepackage{braket}
\usepackage{amsmath}
\usepackage{siunitx}
\usepackage[english]{babel}
\usepackage[utf8]{inputenc}

\usepackage[section]{placeins}

\newcommand{\rcite}{Ref.~\cite}% Prepend \cite with Ref.~
\newcommand{\fref}{Fig.~\ref}

\makeatletter
\let\cat@comma@active\@empty
\makeatother

\begin{document}

\title{Granular aluminum: A superconducting material for high impedance quantum circuits}

\author{Lukas Gr\"unhaupt}
\thanks{Both authors contributed equally.}
\affiliation{Physikalisches~Institut,~Karlsruhe~Institute~of~Technology,~76131~Karlsruhe,~Germany}

\author{Martin Spiecker}
\thanks{Both authors contributed equally.}
\affiliation{Physikalisches~Institut,~Karlsruhe~Institute~of~Technology,~76131~Karlsruhe,~Germany}

\author{Daria Gusenkova}
\affiliation{Physikalisches~Institut,~Karlsruhe~Institute~of~Technology,~76131~Karlsruhe,~Germany}

\author{Nataliya Maleeva}
\affiliation{Physikalisches~Institut,~Karlsruhe~Institute~of~Technology,~76131~Karlsruhe,~Germany}

\author{Sebastian T. Skacel}
\affiliation{Physikalisches~Institut,~Karlsruhe~Institute~of~Technology,~76131~Karlsruhe,~Germany}
\affiliation{Institute~of~Nanotechnology,~Karlsruhe~Institute~of~Technology,~76344~Eggenstein-Leopoldshafen,~Germany}

\author{Ivan Takmakov}
\affiliation{Physikalisches~Institut,~Karlsruhe~Institute~of~Technology,~76131~Karlsruhe,~Germany}
\affiliation{Institute~of~Nanotechnology,~Karlsruhe~Institute~of~Technology,~76344~Eggenstein-Leopoldshafen,~Germany}
\affiliation{Russian Quantum Center, National University of Science and Technology MISIS, 119049 Moscow, Russia}

\author{Francesco Valenti}
\affiliation{Physikalisches~Institut,~Karlsruhe~Institute~of~Technology,~76131~Karlsruhe,~Germany}
\affiliation{Institute~for~Data~Processing~and~Electronics,~Karlsruhe~Institute~of~Technology,~76344~Eggenstein-Leopoldshafen,~Germany}

\author{Patrick Winkel}
\affiliation{Physikalisches~Institut,~Karlsruhe~Institute~of~Technology,~76131~Karlsruhe,~Germany}

\author{Hannes Rotzinger}
\affiliation{Physikalisches~Institut,~Karlsruhe~Institute~of~Technology,~76131~Karlsruhe,~Germany}

\author{Alexey V. Ustinov}
\affiliation{Physikalisches~Institut,~Karlsruhe~Institute~of~Technology,~76131~Karlsruhe,~Germany}
\affiliation{Russian Quantum Center, National University of Science and Technology MISIS, 119049 Moscow, Russia}

\author{Ioan M. Pop}
\email{ioan.pop@kit.edu}
\affiliation{Physikalisches~Institut,~Karlsruhe~Institute~of~Technology,~76131~Karlsruhe,~Germany}
\affiliation{Institute~of~Nanotechnology,~Karlsruhe~Institute~of~Technology,~76344~Eggenstein-Leopoldshafen,~Germany}

\date{\today}

%-----------------------------------------------------------------------------

\begin{abstract}
Superconducting quantum information processing machines are predominantly based on microwave circuits with relatively low characteristic impedance, of about 100 Ohm, and small anharmonicity, which can limit their coherence and logic gate fidelity. 
A promising alternative are circuits based on so-called superinductors, with characteristic impedances exceeding the resistance quantum $R_{\mathrm{Q}} = \SI{6.4}{\kilo\ohm}$. 
However, previous implementations of superinductors, consisting of mesoscopic Josephson junction arrays, can introduce unintended nonlinearity or parasitic resonant modes in the qubit vicinity, degrading its coherence.
Here we present a fluxonium qubit design using a granular aluminum (grAl) superinductor strip.
Granular aluminum is a particularly attractive material, as it self-assembles into an effective junction array with a remarkably high kinetic inductance, and its fabrication can be in-situ integrated with standard aluminum circuit processing.
The measured qubit coherence time $T_2^\mathrm{R}$ up to $\SI{30}{\micro\second}$ illustrates the potential of grAl for applications ranging from protected qubit designs to quantum limited amplifiers and detectors.
\end{abstract}

\maketitle

%%%%%%%%%%%%%%%%%%%%%%%%%%%%%%%%%%%%%%%%%%%%%%%%%%%%%%%%%%%%%%%%%%%%%%%%%%%%%%%%%
%									 Introduction 								%
%%%%%%%%%%%%%%%%%%%%%%%%%%%%%%%%%%%%%%%%%%%%%%%%%%%%%%%%%%%%%%%%%%%%%%%%%%%%%%%%%
Building large scale quantum information processing machines using superconducting circuits \cite{Devoret2013, Gu2017} remains a challenging physics and engineering endeavor. 
Although there are promising small-scale prototypes \cite{Barends2016, Roushan2016, Kandala2017,Langford2017, Otterbach2017, King2018} and proof of principle demonstrations for the necessary building blocks, such as error corrected qubits \cite{Ofek2016, Ghosh2018} or remote entanglement protocols \cite{Roch2014, Narla2016, Kurpiers2018, Dickel2018}, scaling up to large numbers of logical qubits will require breakthroughs in all aspects of qubit technology, including qubit architecture and materials. 
As an example, one of the major challenges facing prevalent transmon qubit \cite{Koch2007} processors is the problem of quantum state leakage towards non-computational degrees of freedom \cite{Willsch2017}, which could become a roadblock for scaling. 
The limited anharmonicity of the transmon may be insufficient to isolate in frequency the qubit computational space from the surrounding, increasingly complex, microwave environment.

A promising alternative qubit architecture is based on so-called superinductors, with characteristic impedance larger than $R_{\mathrm{Q}} = h/(2e)^2 = \SI{6.4}{\kilo\ohm}$, such as the fluxonium qubit \cite{manucharyan_fluxonium_2009}, which offers orders of magnitude larger anharmonicity, and coherence comparable to transmon qubits \cite{pop_coherent_2014}.
In these circuits, quantum fluctuations of the phase dominate over charge fluctuations, and provide a playground for the design of new, potentially protected quantum circuits \cite{Gladchenko2008, Brooks2013, Cohen2017, Puri2017, Sete2017, vool_driving_2018, Groszkowski2018}. Large inductors could also become a building block of next generation flux and phase qubits \cite{Chiorescu2003, Lecocq2012, Yan2016}.
Moreover, microwave resonators employing superinductors and small capacitors have recently been used to boost and confine voltage fluctuations, enabling strong coupling between photons and the electron orbitals in gate-defined quantum dots \cite{Stockklauser2017, Samkharadze2018, Landig2018}. They could also be used in the near future to achieve strong coupling between photons and magnetic moments \cite{Viennot2015}.
Last but not least, high characteristic impedance elements can be a resource for building tailored environments for transport measurements \cite{Corlevi2006, Weissl2015, Arndt2018, martinez2018}. 

However, superinductors are not easy to obtain. One typically relies on the kinetic inductance of an array of tens to hundreds of Josephson junctions \cite{manucharyan_fluxonium_2009, Bell2012, Masluk2012, Stockklauser2017}, patterned in a compact geometry, to reduce stray capacitance. 
Although there are encouraging results \cite{pop_coherent_2014, earnest_realization_2018, lin_demonstration_2018}, the experimental progress has been slowed down by the complexity in fabrication and microwave design required for junction based superinductors. 

In this Letter we present a significantly simplified superinductor design based on an emerging material in the quantum circuits community: granular aluminum (grAl). 
To prove that a grAl superinductor operates as designed, and to quantify its quantum coherence, we use it to build a fluxonium qubit, which we operate and characterize using the tools of circuit quantum electrodynamics (cQED) \cite{Wallraff2004}.
The observed qubit spectrum is in agreement with the one expected from numerical diagonalization of the system Hamiltonian \cite{smith_fluxonium_2016}.
The measured quantum state coherence time  $T_2^\mathrm{R}$ up to \SI{30}{\micro\second} recommends grAl as a competitive alternative to superinductors implemented with mesoscopic Josephson junction arrays, or thin films from other disordered superconductors, like NbN \cite{niepce_high_2018}, NbTiN \cite{Samkharadze2018, Landig2018, hazard_nanowire_2018}, or TiN \cite{peltonen_hybrid_2018, shearrow_atomic_2018}.

Granular aluminum is compatible and convenient to use with current Josephson junction fabrication technology \cite{niemeyer_einfache_1974, dolan_offset_1977} used for pure aluminum circuits, because grAl simply self-assembles when depositing aluminum in a controlled oxygen atmosphere. 
Depending on the oxygen partial pressure, the deposition rate, and substrate temperature, superconducting films with resistivities ranging from $10$ to $10^4$ \SI{}{\micro\ohm\,\centi\meter} and critical temperatures of up to \SI{3.15}{\kelvin} can be achieved \cite{deutscher_transition_1973, pracht_enhanced_2016}.
The kinetic inductance of grAl strips is proportional to their normal state resistance. 
It can therefore be tuned over a wide range,  reaching values up to $\sim$ nH/$\Box$, while maintaining internal quality factors on the order of $10^5$ in the single photon regime \cite{sun_2012, rotzinger_aluminium-oxide_2017, grunhaupt_loss_2018}.
Moreover, the Kerr non-linearity can be reduced by orders of magnitude \cite{maleeva_circuit_2018} compared to superinductors made of mesoscopic Josephson junction arrays \cite{Weissl2015_Kerr}, which helps to suppress the unwanted coupling between the qubit and the environment.  
  
\begin{figure}[t]
\begin{center}
\includegraphics[scale=1.0]{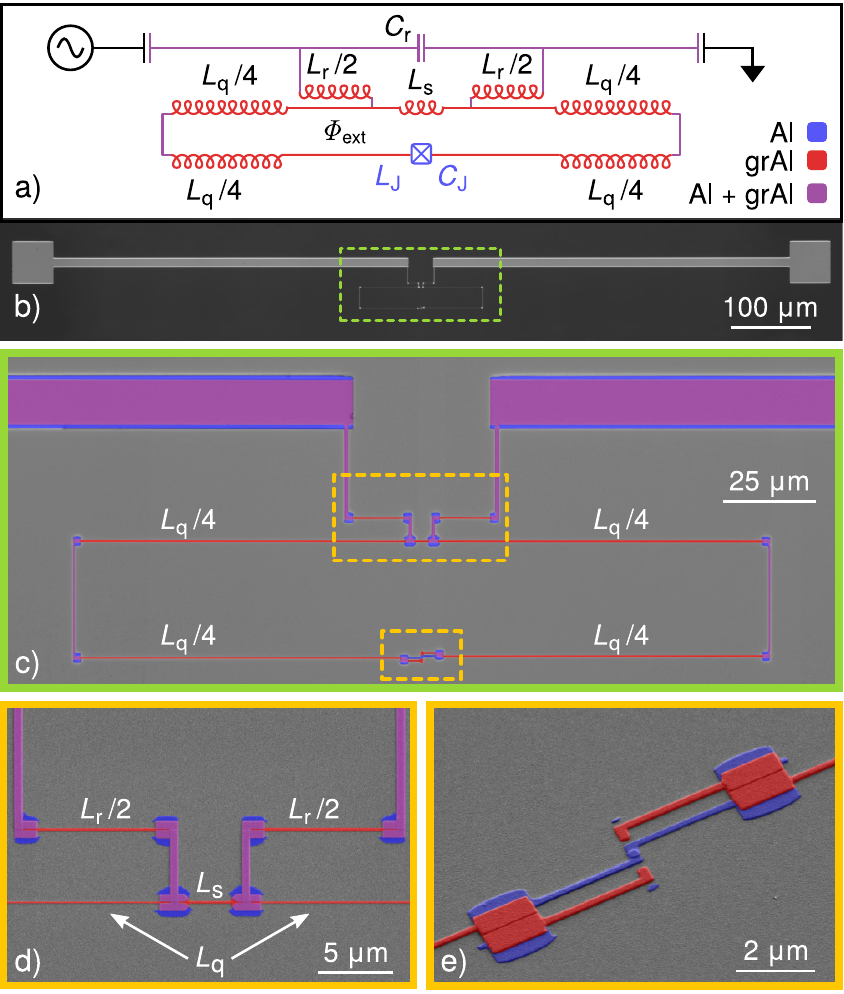}
\caption{\label{fig:1} Fluxonium qubit built using a grAl superinductor. 
\textbf{a)} The qubit consists of a Josephson junction shunted by a \SI{300}{\micro\meter} long grAl superinductor with an estimated characteristic impedance $Z \gtrsim \SI{10}{\kilo\ohm}$ and the first self-resonant mode at \SI{17.4}{\giga\hertz} (see Supplemental Material). 
To perform dispersive readout, we couple the qubit to a microwave resonator through a shared inductance $L_{\mathrm{s}} \approx \SI{1}{\nano\henry}$. The color legend indicates the material used for each circuit element. 
\textbf{b)} Optical microscope image of the readout resonator, which couples to a rectangular waveguide sample holder (not shown) via its dipole moment, following \rcite{kou_simultaneous_2017}. 
\textbf{c)} False colored optical microscope image of the fluxonium qubit inductively coupled to the resonator [cf. green highlighted region in b)]. 
The circuit's inductors are all realized by grAl strips (highlighted in red). 
\textbf{d)} Scanning electron microscope (SEM) image of the resonator-qubit coupling [cf. top orange box in c)]. 
The resonator frequency, coupling strength, and qubit spectrum can all be independently tuned by the length of the corresponding grAl strips. 
\textbf{e)} Tilted SEM image of the fluxonium junction [cf. bottom orange box in c)]. Using a three angle electron beam lift-off process, we connect in-situ a conventional Al/AlOx/Al Josephson junction (highlighted in blue) to the grAl superinductor (see Supplemental Material).
We estimate the junction area $A_\mathrm{J} \approx \SI{0.06}{\micro\meter\squared}$.}
\end{center}
\end{figure}

Figure~\ref{fig:1}a) shows the equivalent circuit of the fluxonium qubit coupled to the readout resonator. 
Shunting a Josephson junction ($L_\mathrm{J}$, $C_\mathrm{J}$) with a large, linear inductor ($L_\mathrm{q}$), leads to an offset charge insensitive superconducting qubit, with a transition frequency tunable over several GHz, and comparable anharmonicity \cite{manucharyan_fluxonium_2009}.
By inductively coupling (via $L_\mathrm{s}$) the qubit to a readout resonator ($L_\mathrm{r}$, $C_\mathrm{r}$), we can measure the qubit state through the shift it induces on the resonator frequency \cite{blais_cavity_2004}. 
In \fref{fig:1}b) we show an optical microscope image of the readout resonator. We place the circuit in a copper rectangular waveguide sample holder, to which the resonator couples via its electric dipole moment, following the scheme reported in \rcite{kou_simultaneous_2017}.

We fabricate the entire circuit comprised of readout resonator and fluxonium in a single electron-beam lithography, three-angle evaporation, lift-off process on a double-side polished c-plane sapphire substrate (see Supplemental Material for technical details). By employing the Niemeyer-Dolan technique \cite{niemeyer_einfache_1974, dolan_offset_1977} and an asymmetric undercut \cite{Lecocq2011} we pattern a conventional Al/AlOx/Al Josephson junction and its short connecting wires [cf. blue highlighted regions in \fref{fig:1}e)]. 
Subsequently, without breaking vacuum, we perform the zero angle deposition of the \SI{40}{\nano\meter} thick grAl film of the superinductor, with sheet resistance $R_s \approx \SI{0.2}{\kilo\ohm\per\Box}$ at room temperature.
The corresponding $\SI{0.8e3}{\micro\ohm\,\centi\meter}$ resistivity is one order of magnitude below the values at which grAl undergoes a superconducting to insulating transition \cite{deutscher_transition_1973}. 
%This value could be increased by several times in future designs, however, in order to do so without risking to lower the coherence, the population of non-equilibrium quasiparticles will have %to be significantly reduced\cite{grunhaupt_loss_2018}. 

\begin{figure*}[t]
\begin{center}
\includegraphics[scale=1.0]{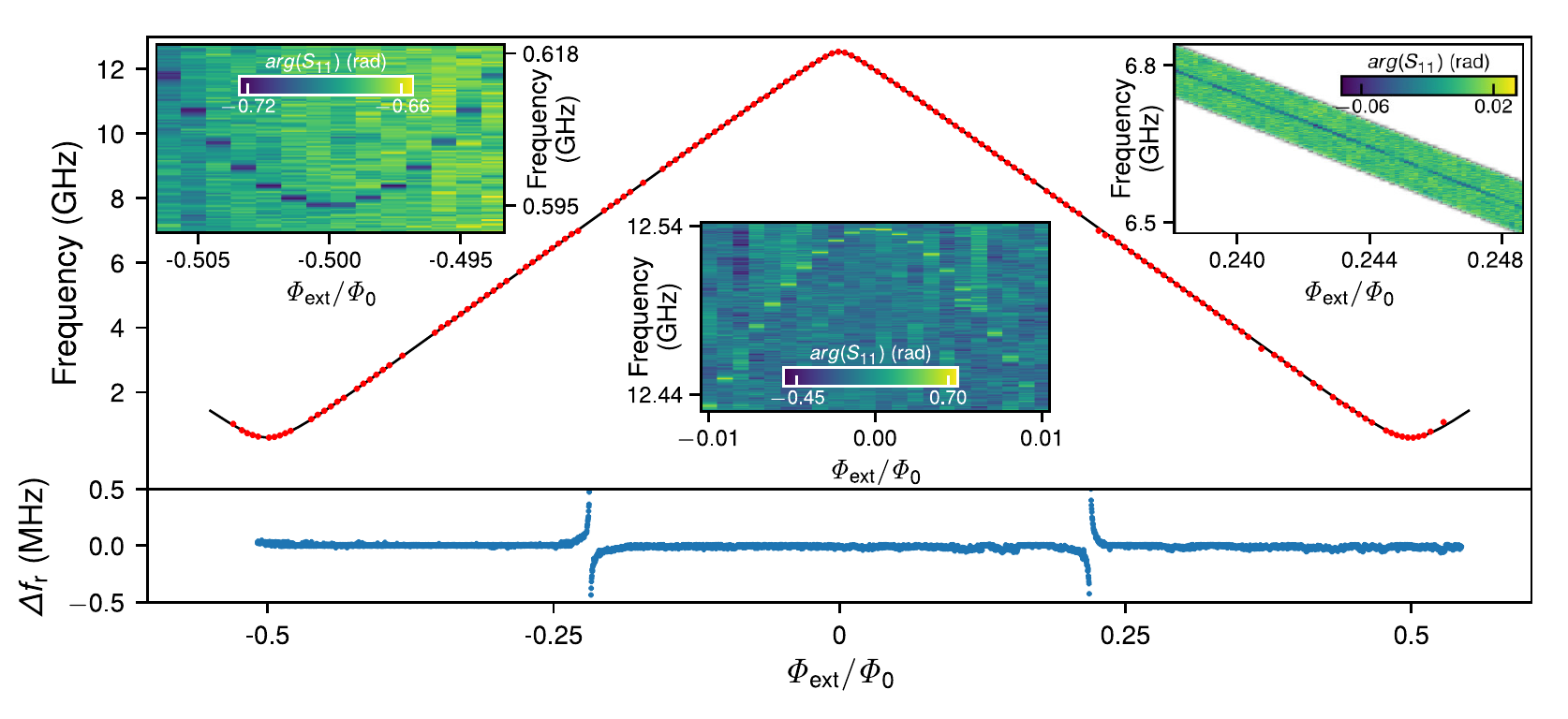}
\caption{\label{fig:2} Fluxonium and readout resonator spectroscopy. Upper panel: Red points show the qubit transition frequency as a function of the external flux measured by two-tone spectroscopy. 
The solid black line is a fit of the fluxonium Hamiltonian following \rcite{smith_fluxonium_2016}. 
We extract a fluxonium inductance $L_\mathrm{q} + L_\mathrm{s} = \SI{225.6}{\nano\henry}$, capacitance $C_\mathrm{J} = \SI{5.2}{\femto\farad}$, and Josephson inductance $L_\mathrm{J} = \SI{13.2}{\nano\henry}$. 
Changing the external magnetic flux $\Phi_\mathrm{ext}$ through the fluxonium loop from half integer to integer flux quanta $\Phi_\mathrm{0}$ tunes the qubit transition frequency from $f^\mathrm{q}_\mathrm{0.5} = \SI{0.594}{\giga\hertz}$ to $f^\mathrm{q}_\mathrm{0} = \SI{12.538}{\giga\hertz}$. 
The top-left and middle insets show zoom-ins into the extremal points of the spectrum at $\Phi_\mathrm{ext}/\Phi_\mathrm{0} = -0.5$ and $\Phi_\mathrm{ext} = 0$, respectively. 
As also visible from the narrowing spectral linewidth, the qubit transition frequency becomes first order flux noise independent at these two symmetry points. 
In the top-right inset we show qubit spectroscopy measurements ranging over \SI{300}{\mega\hertz} for a flux bias close to $\Phi_\mathrm{ext}/\Phi_\mathrm{0}= 0.25$. 
We do not observe anti-crossings, which would be signatures of a strong coupling to two-level defects, one of the mechanisms limiting superconducting qubit coherence \cite{muller_towards_2017}. 
Lower panel: Change of the readout resonator frequency at \SI{7.278}{\giga\hertz} as a function of the external applied flux $\Phi_\mathrm{ext}$ through the fluxonium loop. 
As the qubit frequency crosses the readout resonator frequency, we observe an anti-crossing of about \SI{3}{\mega\hertz}.}
\end{center}
\end{figure*}

The hermetically sealed rectangular waveguide sample holder, anchored to the $25$~mK stage of a commercial dilution cryostat, is placed inside a copper-aluminum shield covered with infrared absorbing coating and further enclosed by a $\mu$-metal shield (cf. \rcite{grunhaupt_loss_2018}).
%
%Multiple commercial microwave attenuators, distributed over different temperature stages of the cryostat provide 70 dB of attenuation and reduce noise from room temperature and higher cryostat stages. Different commercial and home-made lowpass filters ensure additional shielding from photons with frequencies $ > \SI{8.5}{\giga\hertz}$. 
%In the output line the signal is amplified by 43 dB using a commercial high electron mobility transistor amplifier on the \SI{1.6}{\kelvin} stage of the cryostat and by an additional 63 dB at room temperature.
%
To measure the fluxonium spectrum we perform standard two-tone microwave spectroscopy, measuring the complex reflection coefficient $S_{11}$ of the resonator, while sweeping a second generator in the range of expected qubit frequencies. 
Figure~\ref{fig:2} shows the measured fluxonium spectrum (red points), $\Phi_\mathrm{0}$ periodic as a function of the external flux $\Phi_\mathrm{ext}$. 
%
%The shown fluxonium allows to tune the fundamental $\ket{g}$ to $\ket{e}$ transition over a frequency range $\sim \SI{12}{\giga\hertz}$, from \SI{594}{\mega\hertz} at $\Phi_\mathrm{ext}/\Phi_\mathrm{0} = n/2$ to \SI{12.538}{\giga\hertz} at $\Phi_\mathrm{ext}/\Phi_\mathrm{0} = n$, with $n \in \mathbb{Z}$ . 
%The upper left and bottom middle inset in \fref{fig:2} show zoom-ins into the two extremal points of the spectrum at $\Phi_\mathrm{ext}/\Phi_\mathrm{0} = - 0.5$ and $\Phi_\mathrm{ext}/\Phi_\mathrm{0} = 0$ respectively and reveal the steep frequency dependence of the qubit frequency with external flux. 
%
From a numerical fit (cf. black line in \fref{fig:2}) of the fluxonium Hamiltonian to the measured transition frequencies, we extract the total inductance of the loop $L_\mathrm{q} + L_\mathrm{s} = \SI{225.6}{\nano\henry}$, corresponding to a sheet inductance $L_\mathrm{kin} = \SI{0.1}{\nano\henry\per\Box}$, the junction capacitance $C_\mathrm{J} = \SI{5.2}{\femto\farad}$, and the Josephson inductance $L_\mathrm{J} = \SI{13.2}{\nano\henry}$.
%
%In the top right inset in \fref{fig:2} we show qubit spectroscopy measurements close to $\Phi_\mathrm{ext}/\Phi_\mathrm{0} = 0.25$, corresponding to a qubit transition frequency between \SI{6.5}{\giga\hertz} and \SI{6.8}{\giga\hertz}. 
%We do not observe anti-crossings of the fluxonium with two-level systems (TLS) between \SI{6.2}{\giga\hertz} and \SI{6.8}{\giga\hertz} (for clarity only partially shown in the inset). 
%At the sweep resolution used for the full frequency range spectroscopy, which is a factor of 40 more coarse we do not observe any anti-crossings either. 
%Consequently, we believe that strong, coherent coupling to TLS is not more common in our fabricated design than in fluxonium realizations using mesoscopic Josephson junction chains. 
%However, mostly TLS situated in the oxide barrier of the Josephson junction are subjected to electric fields high enough to achieve strong coherent coupling. 
%Since our design employs a small Josephson junction with an area of $\sim \SI{0.1}{\micro\meter\squared}$, not observing anti-crossings is consistent with measurements in other superconducting qubits. 
%To investigate the distribution of TLS not sufficiently strongly coupled to produce anti-crossings, a detailed study of $T_1$ as a function of qubit frequency would have to be carried out, which is outside the scope of our current manuscript.

The lower panel of \fref{fig:2} shows the shift of the readout resonator frequency vs. externally applied flux $\Phi_\mathrm{ext}$ through the fluxonium loop, relative to the dressed resonator frequency of \SI{7.278}{\giga\hertz}.
We estimate a qubit state dependent dispersive shift of the resonator frequency in the range of $\SI{130}{\kilo\hertz}$ at $\Phi_\mathrm{ext}/\Phi_\mathrm{0}=-0.5$ (cf. color scale in \fref{fig:2} and the resonator linewidth of $\SI{3.2}{\mega\hertz}$).

\begin{figure*}[t]
\begin{center}
\includegraphics[scale=1.0]{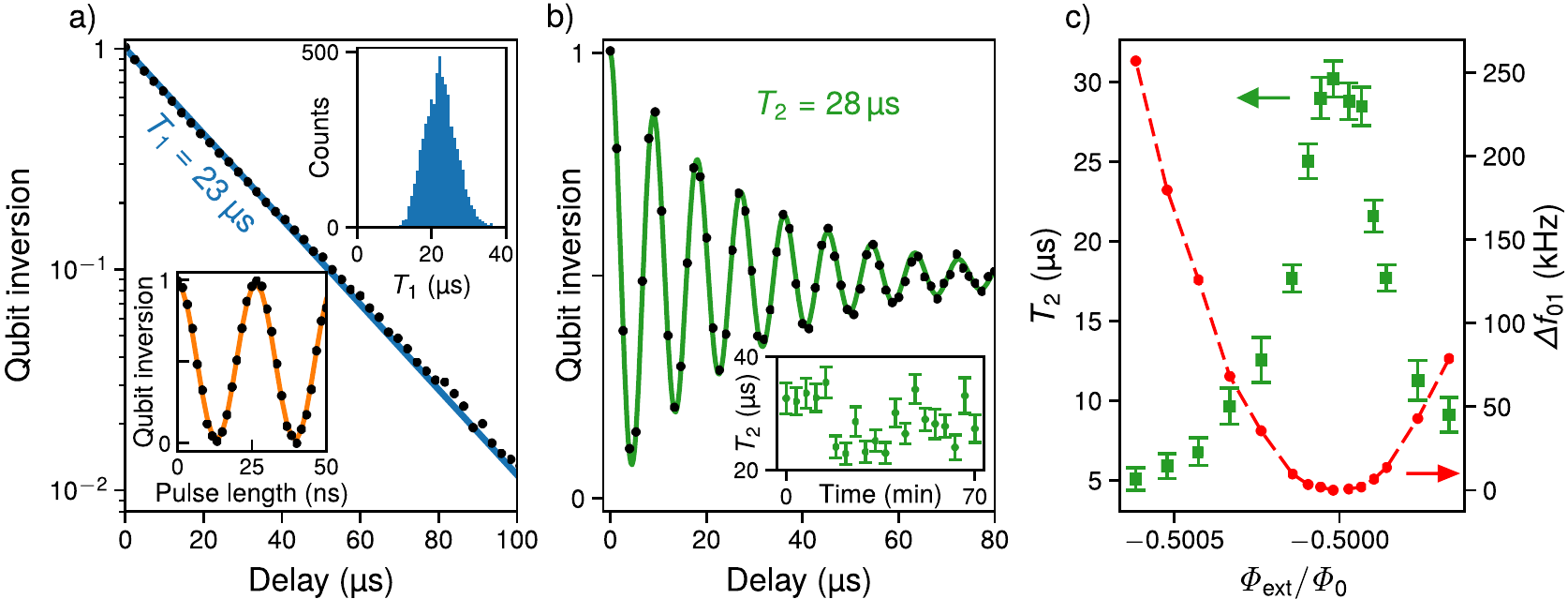}
\caption{\label{fig:3} Quantum coherence of the fluxonium superconducting qubit with grAl superinductor. 
\textbf{a)} Average energy relaxation time $T_1 = \SI{23}{\micro\second}$ at the $\Phi_\mathrm{ext}/\Phi_\mathrm{0} = -0.5$ sweet spot, averaged over $\sim 6000$ measurements taken over a total time of $\sim \SI{17}{\hour}$ (measured data - black points, single exponential fit - blue solid line). 
The histogram in the upper inset shows the distribution of individual $T_1$ measurements. 
Rabi oscillations \cite{rabi_space_1937} of the qubit with a frequency of \SI{38}{\mega\hertz}, limited by the maximum output power of our microwave generator are shown in the lower inset. 
A $\pi$-pulse, inverting the qubit population, corresponds to a square-envelope pulse with a length of \SI{13}{\nano\second}.
\textbf{b)} The black points show the result of a Ramsey fringes measurement \cite{ramsey_molecular_1950} for a \SI{110}{\kilo\hertz} detuned drive with respect to the qubit frequency of $f^\mathrm{q}_\mathrm{0.5} = \SI{594.37}{\mega\hertz}$ at the $\Phi_\mathrm{ext}/\Phi_\mathrm{0} = -0.5$ sweet spot, averaged over \SI{70}{\minute}.
From the fit of an exponentially decaying cosine (green solid line) we extract a coherence time $T_2^\mathrm{R} = \SI{28}{\micro\second}$, comparable to fluxonium qubits with superinductors made from mesoscopic Josephson junction arrays \cite{pop_coherent_2014, lin_demonstration_2018, earnest_realization_2018, vool_driving_2018}. 
In the inset, we show the fluctuations of $T_2^\mathrm{R}$ over time by fitting shorter averaged Ramsey measurements. 
\textbf{c)} Flux dependence of the coherence time $T_2^\mathrm{R}$ (green squares) and the measured qubit detuning with respect to $f^\mathrm{q}_\mathrm{0.5}$ (red points) close to the half integer sweet spot. The red dashed line connecting the points is a guide to the eye.} 
\end{center}
\end{figure*}

To quantify the quantum coherence of the grAl fluxonium qubit we perform standard time-domain manipulations and measurements (see the Supplemental Material for a detailed setup schematic).
Figure~\ref{fig:3} shows the results obtained at the $\Phi_\mathrm{ext}/\Phi_\mathrm{0}=-0.5$ sweet spot, where energy relaxation due to non-equilibrium quasiparticles tunneling through the Josephson junction is suppressed \cite{pop_coherent_2014}, the spectrum is first order flux noise insensitive \cite{Manucharyan2012} and the qubit coherence time shows a maximum [see Fig.~\ref{fig:3}c)]. 
We extract typical energy relaxation times $T_1$ in the range of $\SI{20}{\micro\second}$ to $\SI{30}{\micro\second}$, and Ramsey coherence times $T_2^\mathrm{R}$ up to $\SI{30}{\micro\second}$.

By performing Rabi oscillations \cite{rabi_space_1937} with constant microwave drive power and varying duration, we achieve qubit population inversion with a frequency of \SI{38}{\mega\hertz}, which corresponds to a $\pi$-pulse of $\sim \SI{13}{\nano\second}$ [see lower left inset in \fref{fig:3}a)], orders of magnitude faster than the coherence time.

We measured the relaxation time $T_1$ of the grAl fluxonium repeatedly, performing a total of 6000 measurements over the course of $\sim \SI{17}{\hour}$. 
The averaged measurement result plotted in log-lin scale in \fref{fig:3}a) shows an exponential behavior, and the histogram of the individual measurements is shown in the upper right inset in \fref{fig:3}a). From the distribution we obtain an average $T_1 = \SI[separate-uncertainty=true]{23 \pm 4}{\micro\second}$. The measured $T_1$ values could be limited either by dielectric losses or by non-equilibrium quasiparticles \cite{Palacios-Laloy2009, Riste2013, Wenner2013, Gustavsson2016, Serniak2018} in the grAl superinductor \cite{vool_non-poissonian_2014, grunhaupt_loss_2018}. 
If excess quasiparticles in the grAl superinductor are the limiting loss mechanism at the $\Phi_\mathrm{ext}/\Phi_\mathrm{0}=-0.5$ sweet spot, we extract a normalized density $x_\mathrm{qp} = 4.1 \times 10^{-7}$ (following the methodology in Ref.~\cite{pop_coherent_2014}). This value is two orders of magnitude higher than previously observed in a Josephson junction array superinductor \cite{pop_coherent_2014}, but also one order of magnitude lower than values measured in grAl resonators \cite{grunhaupt_loss_2018}. 
The quasiparticle density in grAl is presumably increased compared to pure aluminum, due to the longer quasiparticle lifetime \cite{grunhaupt_loss_2018}. 

%
%Figure~\ref{fig:3}b shows the result of a Ramsey fringes \cite{ramsey_molecular_1950} measurement at the $\Phi_\mathrm{ext}/\Phi_\mathrm{0} = -0.5$ sweet spot averaged over \SI{70}{\minute}.
%We detune the microwave drive tone used to manipulate the qubit by \SI{110}{\kilo\hertz} and observe coherent oscillations of the qubit population up to \SI{80}{\micro\second}. 
%From the fit of an exponentially damped cosine (green solid line) we extract a quantum coherence time $T_2 = \SI{28}{\micro\second}$. 

A spin-echo measurement performed by introducing a $\pi$-pulse in the middle of the Ramsey sequence increases $T_2^\mathrm{E}$ up to \SI{46}{\micro\second}, close to the theoretical limit of $T_2 = 2 T_1$. 
We extract a dephasing time, dominated by low frequency noise, $ T_\phi = 2 T_1 T_2^\mathrm{R} / (2 T_1 - T_2^\mathrm{R}) \approx \SI{72}{\micro\second}$.    
The fluctuations of $T_2^\mathrm{R}$ can be due to residual magnetic flux noise, adsorbed surface spins \cite{Kumar2016}, or the effect of fluctuating non-equilibrium quasiparticle numbers \cite{Catelani2012}.

In the region in-between the flux sweet spots, where the frequency of the qubit is strongly susceptible to flux noise, $T_{2}$ is reduced to values in the range of \SI{50}{\nano\second}, which could be explained by residual flux noise. The corresponding flux noise amplitude $A = \SI{30}{\micro\Phi_0}$ is about a factor of three larger than observed in devices using Josephson junction superinductors \cite{Kou_fluxonium_2017}, and it might be due to the longer superinductor loop. 

At the zero flux sweet spot, the measured energy decay is non-exponential (see Supplemental Material). It can be fitted to a model assuming a residual decay rate $\Gamma_{R} =\SI{1/13}{\micro\second}^{-1}$ and an additional decay $\Gamma_\mathrm{qp} = \SI{1/3.2}{\micro\second}^{-1}$ due to the presence of quasiparticles tunneling across the Josephson junction. From the measured $\Gamma_\mathrm{qp}$ we calculate an excess quasiparticle density $x_\mathrm{qp} = 1.2 \times 10^{-5}$ in the vicinity of the junction.  
This excess quasiparticle population can be explained by the larger superconducting gap of grAl (\SI{300}{\micro\eV}) compared to thin film aluminum (\SI{230}{\micro\eV}), which effectively traps quasiparticles. 
The coherence time $T_{2}^\mathrm{R}=\SI{3.8}{\micro\second}$ at the zero flux sweet spot is also reduced, as expected from the measured excess quasiparticle population \cite{Catelani2012}.

In summary, we have demonstrated that granular aluminum is a viable material for the implementation of superinductors, and its deposition can be successfully integrated in the fabrication process used for Josephson junctions with pure aluminum electrodes. 
The measured grAl fluxonium qubit shows state-of-the-art coherence times in the range of tens of $\SI{}{\micro\second}$, while the gate operation time can be as short as a few $\SI{}{\nano\second}$.
If necessary, the currently reported value of the characteristic kinetic inductance of the grAl film, $L_{\mathrm{kin}} = \SI{0.1}{\nano\henry\per\Box}$, can be increased by an order of magnitude by using a thinner and stronger oxidized grAl film.
Despite the disordered nature of grAl, a material that incorporates a significant amount of amorphous aluminum oxide, from spectroscopy and time domain measurements we conclude that grAl is a suitable material for superconducting quantum hardware.

We believe that grAl superinductors will enable the realization of increasingly complex, and potentially protected qubit designs. Similarly to the fluxonium qubit, which confines its electromagnetic fields to the parallel plate capacitor of the Josephson junction, and to the internal degrees of freedom of the kinetic inductor, the emerging electronics will be less vulnerable to cross-talk and radiation loss. This could open a new technological avenue towards the up-scaling of quantum coherent superconducting circuits.

%% Acknowledgement %%%
%\begin{acknowledgments}
We are grateful to A. Bilmes, J. Lisenfeld, C. Smith, W. Wernsdorfer, and M. Wildermuth for insightful discussions, and to J. Ferrero, A. Lukashenko, and L. Radtke for technical assistance. 
Funding was provided by the Alexander von Humboldt foundation in the framework of a Sofja Kovalevskaja award endowed by the German Federal Ministry of Education and Research, and by the Initiative and Networking Fund of the Helmholtz Association, within the Helmholtz Future Project \textit{Scalable solid state quantum computing}. 
This work has been partially supported by the European Research Council advanced grant MoQuOS (N. 741276). 
I.T. and A.V.U. acknowledge partial support from the Ministry of Education and Science of the Russian Federation in the framework of the Increase Competitiveness Program of the National University of Science and Technology MISIS (Contract No. K2-2017-081).
Facilities use was supported by the KIT Nanostructure Service Laboratory (NSL). We acknowledge qKit \cite{qkit} for providing a convenient measurement software framework.
%\end{acknowledgments}

\bibliography{2018_grAl_fluxonium}

\onecolumngrid
\renewcommand{\thefigure}{S\arabic{figure}}
\renewcommand{\thetable}{S\arabic{table}}
\renewcommand{\theequation}{S\arabic{equation}}

\setcounter{figure}{0}
\setcounter{table}{0}

\section*{Supplemental material}

\subsection*{Coherence times at $\Phi_\mathrm{ext} = 0$}

\begin{figure}[h]
\begin{center}
\includegraphics[scale=1.0]{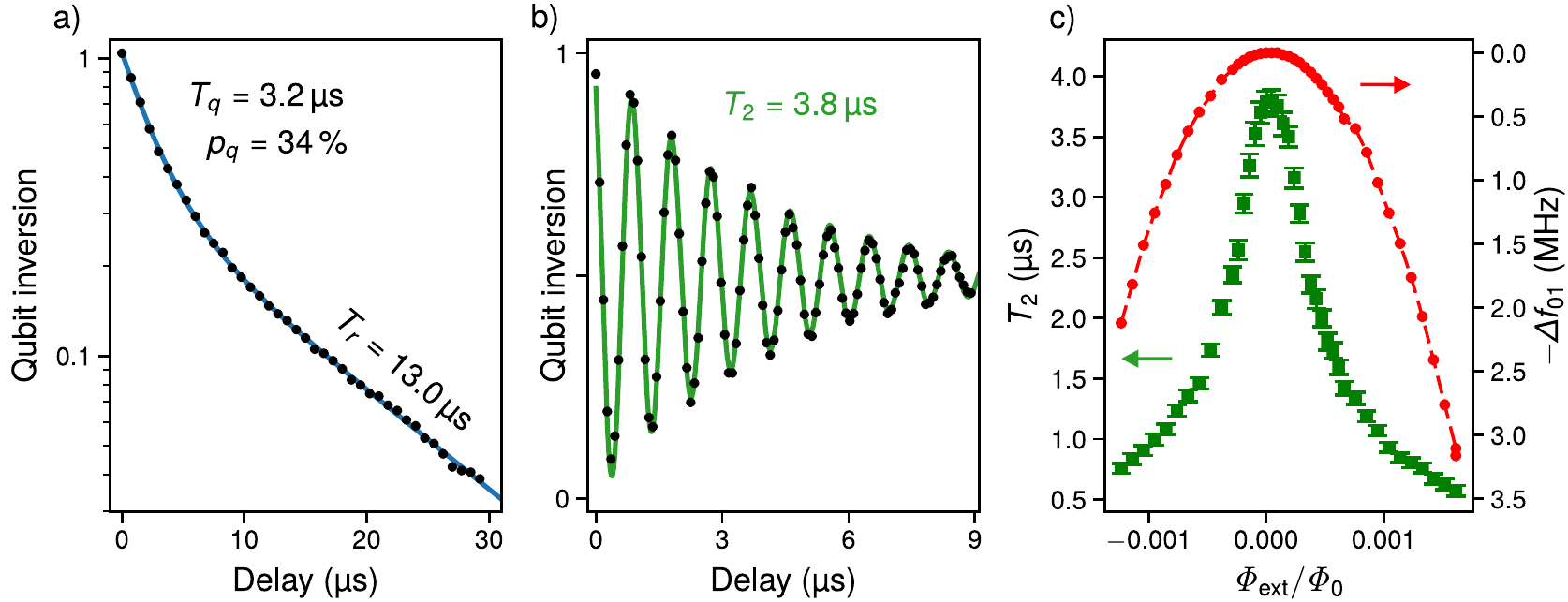}
\caption{a) Relaxation time $T_1$ at $\Phi_\text{ext} = 0$. The relaxation is well described by a sum of two exponential functions (see Eq.~\eqref{SumExpo}), assuming time traces with a remaining relaxation time $T_r$ and traces with increased relaxation, at a probability $p_q$, giving rise to an additional relaxation $T_q$ (measured data - black points, fit - blue solid line) b) The black points show the result of a Ramsey fringes measurement for a \SI{1.15}{MHz} detuned drive with respect to the qubit frequency of $f^q_{0.0} = \SI{12.540}{\giga\hertz}$ at $\Phi_\text{ext}  = 0$, averaged over \SI{30}{min}. From the fit of an exponentially decaying cosine (green solid line) we extract a coherence time $T_2^\mathrm{R} = \SI{3.8}{\micro\second}$.
c) Flux dependence of the coherence time $T_2^\mathrm{R}$ (green squares) and the measured qubit detuning (red points) with respect to $f^q_{0.0}$ close to zero flux. The red dashed line connecting the points is a guide to the eye.} \label{Fig4}
\end{center}
\end{figure}

The dynamic of the $T_1$ relaxation shown in Fig.~\ref{Fig4}a is well described by a sum of two exponential decays. This might be explained by the existence of an additional decay channel, on average present with a probability $p_q$. The timescale over which this channel influences the energy relaxation needs to be longer than the measurement pulse sequence, which here is \SI{30}{\micro\second}. Thus we average measurements with no additional decay resulting in $T_r$ and traces with an additional decay $T_q$. We fit the measured $T_1$ relaxation with

\begin{equation}
P(t) = p_q \cdot e^{- \left(\frac{1}{T_q} + \frac{1}{T_r} \right) \cdot t} + (1 - p_q) \cdot e^{- \frac{1}{T_r} \cdot t}. \label{SumExpo}
\end{equation}

We attribute this additional decay channel to quasiparticles tunneling across the Josephson junction, similar to \cite{pop_coherent_2014}.

At the $\Phi_\mathrm{ext}/\Phi_0 = 0$ sweet spot we measure a coherence time $T_2^\mathrm{R} = \SI{3.8}{\micro\second}$ using a Ramsey fringes measurement \cite{ramsey_molecular_1950}, about one order of magnitude shorter than at the $\Phi_\mathrm{ext}/\Phi_0 = 0.5$ sweet spot (cf. Fig.~3 in the main text). 
Figure~\ref{Fig4}c shows $T_2^\mathrm{R}$ (green squares) and the qubit transition frequency relative to the maximum value of $f^q_{0.0} = \SI{12.540}{\giga\hertz}$ (red circles) as a function of the externally applied magnetic flux. 

\subsection*{Sample fabrication}

\begin{figure}[h]
\begin{center}
\includegraphics[scale=.7]{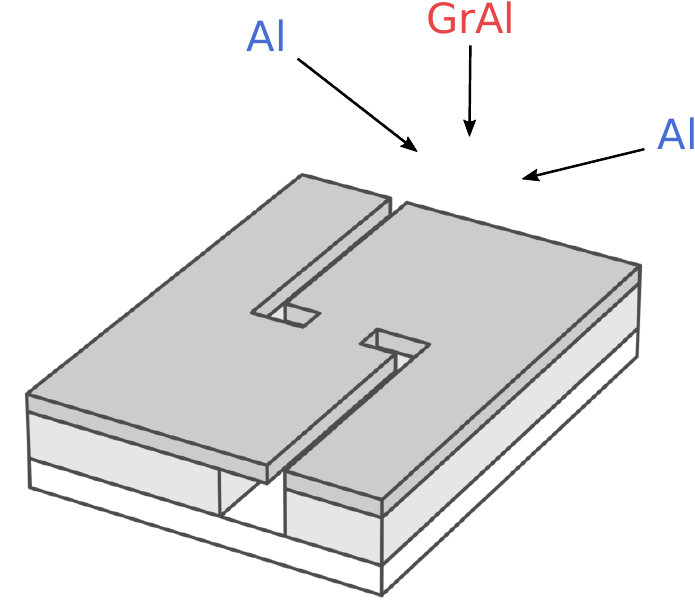}
\caption{\label{Fig6} Sketch of the resist stack for structuring the Josephson junction. It combines a Niemeyer-Dolan bridge \cite{niemeyer_einfache_1974, dolan_offset_1977} in its middle with an asymmetric undercut \cite{Lecocq2011} for the feedlines, which are patterned by a double-angle evaporation of pure aluminum.
The final zero angle evaporation of grAl does not cover the aluminum film in the vicinity of the Josephson junction and its feedlines. A connection between the junction and the superinductor is ensured by connection pads at the end of the $\sim \SI{2}{\micro\meter}$ long feedlines (not shown, cf. Fig.~1 in the main text).} 
\end{center}
\end{figure}

We fabricated the entire circuit by a three-angle evaporation process using a PMMA/(PMMA-MMA) resist stack on a double-side polished c-plane sapphire substrate. Fig.~\ref{Fig6} schematically shows the lithography mask for the Al/AlOx/Al Josephson junction, which is patterned using a \SI{50}{\kilo\electronvolt} $e$-beam writer. The mask combines a Niemeyer-Dolan bridge \cite{niemeyer_einfache_1974, dolan_offset_1977} with an asymmetric undercut \cite{Lecocq2011} for the feedlines. In a first step the Josephson junction is patterned by a two-angle aluminum evaporation process with a thickness of \SI{20}{\nano\meter} and \SI{30}{\nano\meter}, respectively. Thereby, all wires of the design parallel to the evaporation direction, and the antenna, due to its width, are deposited (cf. Fig.~1 in the main text).
Finally, without breaking the vacuum, we patterned all inductive parts of the circuit by a zero-angle evaporation of a \SI{40}{\nano\meter} thick grAl film with a resistivity $\rho = \SI{0.8e3}{\micro\ohm\centi\meter}$.
The junction and its feedlines (see Fig.~\ref{Fig6} and cf. Fig.~1 in the main text) are connected to the grAl film using connection pads $\sim \SI{2}{\micro\meter}$ away from the Josephson junction.

\subsection*{Time-domain setup}

\begin{figure}[h]
\begin{center}
\includegraphics[scale=.9]{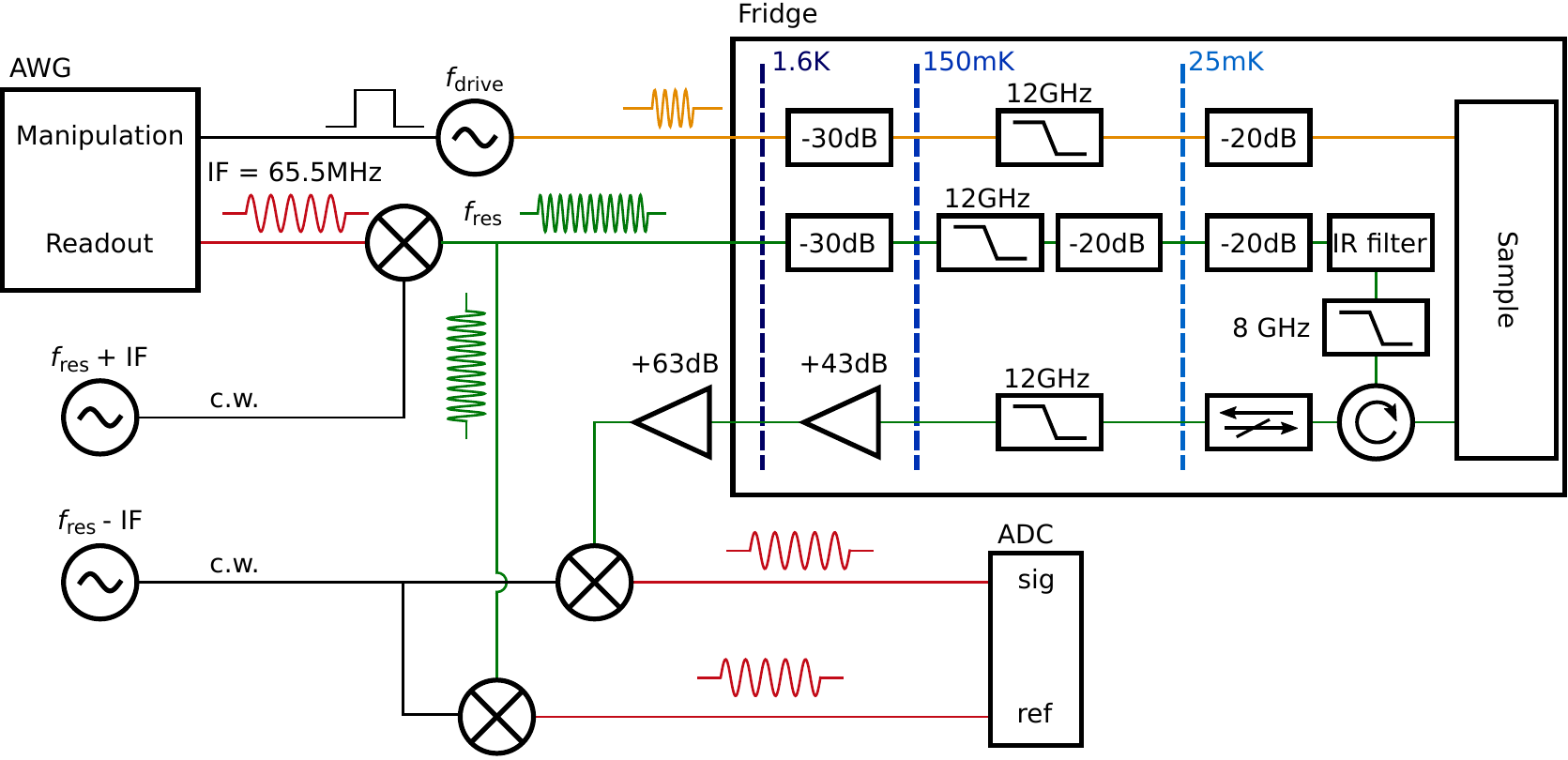}
\caption{Interferometer setup for time-domain manipulation and measurements. Two separate channels of a commercial arbitrary waveform generator (AWG) provide the pulse shaping abilities for the readout and manipulation pulses. We perform homodyne manipulation by feeding the square envelope pulse from the AWG into the $IQ$-modulation ports of a commercial microwave vector generator set to the qubit transition frequency. The readout uses a low-cost commercial two-channel microwave generator operated in continuous wave mode in addition to an AWG channel in an interferometric configuration. First, the readout pulse at an intermediate frequency IF = \SI{65.5}{\mega\hertz} is upconverted to the readout frequency and split into two signals. One signal is directly measured with the analog digital converter (ADC), whereas the other signal first passes the fridge. Both signals are interfered computationally to extract the $I$ and $Q$ quadratures. All microwave lines in the cryostat are attenuated and filtered using commercially available components. An additional home-made infrared (IR) filter employing Stycast\textsuperscript{\textregistered}, and designed to have an impedance of \SI{50}{\ohm} at cryogenic temperatures, ensures an attenuation of more than $\SI{-10}{\decibel}$ for frequencies larger than \SI{60}{\giga\hertz}.}
\end{center}
\end{figure}

\subsection*{Characteristic impedance of the superinductor}

From fitting the measured fluxonium spectrum, we obtain a total superinductance of $L_\mathrm{total} = \SI{225.6}{\nano\henry}$ and a fluxonium capacitance $C_\mathrm{J} = \SI{5.2}{\femto\farad}$ (see main text), which leads to a qubit plasmon mode impedance of $Z = \SI{6.6}{\kilo\ohm}$. 
In order to estimate the impedance of the bare superinductor, we measured the Junction size using a SEM image (cf. Fig. 1e in the main text) to be \SI{0.06}{\micro\meter\squared} with an error of $20 \%$. Using a Josephson junction capacitance per area of $c_\mathrm{J} = \SI{50}{\femto\farad\per\micro\meter\squared}$ we calculate a capacitance of \SI{3}{\femto\farad} for the fluxonium junction alone. This yields a capacitance $C_\mathrm{s} = \SI{2.2}{\femto\farad}$ associated with the superinductor. 
Using these values we obtain a superinductor characteristic impedance $Z~=~\sqrt{L_\mathrm{total}/C_\mathrm{s}}~\approx~\SI{10}{\kilo\ohm}$.

For comparison we simulated the superinductor loop, approximating the junction with an ideal capacitor with \SI{3}{\femto\farad}. 
We obtain a resonant frequency of \SI{5.4}{\giga\hertz} (corresponding to the qubit plasmon mode), from which we calculate a superinductor self capacitance of $C'_\mathrm{s} = \SI{0.9}{\femto\farad}$. Using this method, the superinductor characteristic impedance is estimated to be $Z = \SI{16}{\kilo\ohm}$.
In this simulation, we find the next self-resonant mode of the superinductor at \SI{17.4}{\giga\hertz}, well above the qubit spectrum.

\end{document}